\documentclass{siamltex}

\newcommand{\MyGamma}[1]{\overline{#1}}
\newcommand{\GG}[0]{{\cal G}}
\newcommand{\HH}[0]{{\cal H}}
\newcommand{\TT}[0]{{\cal T}}

\newtheorem{problem}{Problem}
\newtheorem{fact}{Fact}
\newtheorem{claim}{Claim}

\newcommand{\CiteAllStat}[0]{%
\cite{BCS83,
Cox75,
Cox77,
Cox78,
Cox80,
CS79,
Denning82, 
Sande77,
Sande78,
Sande84,
Sande?}}

\begin{document}

\newcommand{\kaoproblem}[1]{
\begin{problem}\rm
#1
\end{problem}}

\newcommand{\namedkaoproblem}[2]{
\begin{problem}[#1]\rm
#2
\end{problem}}

\title{Data Security Equals Graph Connectivity\thanks{A preliminary
version of this work appeared in Proc.~2nd International Workshop on
Discrete Mathematics and Algorithms, Guanzhou, China, December 18--20,
1994, pp.~134--147.}}

\author{Ming-Yang Kao\thanks{Department of Computer Science,
Duke University, Durham, NC 27708. Supported in part by NSF
Grants MCS-8116678, DCR-8405478, and CCR-9101385. Part of this
work was done while the author was at the Department of Computer
Science, Yale University, New Haven, Connecticut 06520.}}

\maketitle

\begin{abstract}
To protect sensitive information in a cross tabulated table, it
is a common practice to suppress some of the cells in the table.
This paper investigates four levels of data security of a
two-dimensional table concerning the effectiveness of this
practice.  These four levels of data security protect the
information contained in, respectively, individual cells,
individual rows and columns, several rows or columns as a whole,
and a table as a whole.  The paper presents efficient algorithms
and NP-completeness results for testing and achieving these four
levels of data security.  All these complexity results are
obtained by means of fundamental equivalences between the four
levels of data security of a table and four types of connectivity
of a graph constructed from that table.
\end{abstract}



\section{Introduction}\label{sec_intro}
Cross tabulated tables are used in a wide variety of documents to
organize and exhibit information.  The values of sensitive cells
in such tables are routinely suppressed to conceal sensitive
information.  There are two fundamental issues concerning the
effectiveness of this practice \CiteAllStat.  One is whether an
adversary can deduce significant information about the suppressed
cells from the published data of a table.  The other is how a
table maker can suppress a small number of cells in addition to
the sensitive ones so that the resulting table does not leak
significant information.

This paper investigates how to protect the information in a
two-dimensional table that publishes three types of data (see
{\cite{KaoG93}} for examples): (1) the values of all cells except a
set of sensitive ones, which are {\it suppressed}, (2) an upper
bound and a lower bound for each cell, and (3) all row sums and
column sums of the complete set of cells.  The cells may have
real or integer values. They may have different bounds, and the
bounds may be finite or infinite.  The upper bound of a cell
should be strictly greater than its lower bound; otherwise, the
value of that cell is immediately known even if that cell is
suppressed.  The cells that are not suppressed also have upper
and lower bounds. These bounds are necessary because some of the
unsuppressed cells may later be suppressed to protect the
information in the sensitive cells.

The focus of this paper is on how to protect the type of
information defined here.  A {\it bounded feasible assignment} to
a table is an assignment of values to the suppressed cells such
that each row or column adds up to its published sum and the
bounds of the suppressed cells are all satisfied.  A linear
combination of the suppressed cells is a {\it linear invariant}
if it has the same value at all bounded feasible assignments (see
{\cite{KaoG93}} for examples).  Intuitively, the information
contained in a linear invariant is unprotected because its value
can be uniquely deduced from the published data.  Five classes of
linear invariants are of special significance.  A {\it positive}
invariant is one whose coefficients are all nonnegative with at
least one coefficient being positive.  A {\it unitary} invariant
is one whose coefficients are $+1$, $0$, or $-1$.  A {\it sum}
invariant is one whose coefficients are $+1$ or $0$. A {\it
rectangular} sum invariant is one that sums over all suppressed
cells shared by a set of rows and a set of columns.  An {\it
invariant cell} is a suppressed cell that forms a linear
invariant all by itself.

Four levels of data security of a table are discussed in this
paper.  To motivate the discussion, suppose that a given table
tabulates the quantities of several products made by different
factories. A row represents a factory, a column records the
quantities of a product, and a cell contains the quantity of a
product made by a factory.
Level 1 protects the suppressed cells individually.  A factory
wishes to conceal the quantity of a particular product.
Naturally, that quantity should be suppressed and its precise
value should not be uniquely determined from the published data
of the table.  Thus, a suppressed cell is {\it protected} if it
is not an invariant cell {\cite{Gusfield88}}.
Level 2 protects a row (or column) as a whole.  After all
suppressed cells are protected, an adversary may still be able to
obtain useful information by combining the suppressed cells.  If
a factory wishes to protect the information about the quantities
of all its products as a whole, it must ensure that no
information of a sensitive type can be extracted by combining the
suppressed cells in the row representing that factory.  Hence, a
row is {\it protected} if there is no linear invariant of a
desired type that combines the suppressed cells in that row.
Level 3 protects a set of $k$ rows (or $k$ columns) as a whole.
Suppose that a company owns $k$ factories. It wishes to conceal
aggregate information about all its factories, not just the
information about each individual factory.  It should require
that no information of an important type may be derived by
combining the suppressed cells in the $k$ rows for its $k$
factories. Thus a set of $k$ rows is {\it protected} if there is
no linear invariant of a desired type that combines the
suppressed cells in those rows.
Level 4 protects the given table as a whole.  Further suppose
that the above company owns all the factories tabulated in the
table.  It wishes to protect aggregate information about of all
its factories and all their products.  It stipulates that only
trivial information may be found among a desired class of
combinations of the suppressed cells. Thus, a table is {\it
protected} if it has no linear invariant of a desired type that
combines its suppressed cells.

The key contribution of this paper is to establish that the
latter three levels of data security of a table are equivalent to
three types of connectivity of a graph called the {\it suppressed
graph} of that table. Previously, Gusfield showed that the first
level of data security is equivalent to a certain type of
connectivity of the suppressed graph {\cite{Gusfield88}}.  The
paper further uses these fundamental equivalences to obtain three
sets of complexity results.  Firstly, the second and the fourth
level of data security of a table can be tested in optimal linear
time and that the third level can be tested in polynomial time.
Previously, Gusfield showed how to find all invariant cells of a
table and test for its first level of data security in optimal
linear time {\cite{Gusfield88}}.  Secondly, for each of the four
levels of data security it is an NP-complete problem to compute
and suppress the minimum number of additional cells in a table in
order to achieve the desired level of data security.  Thirdly,
for a large and practical class of tables, the above optimal
suppression problem for the second and the fourth level of data
security can be solved in optimal linear time.  For the first
level of data security, Gusfield showed that the optimal
suppression problem can be solved in optimal linear time
{\cite{Gusfield87}}.  For the third level of data security the
optimal suppression problem remains open.

We review basics of graphs and tables
in~\S\ref{sec_basics_graph}, discuss the four levels of data
security in {\S\ref{sec_cell}} through {\S\ref{sec_table}}, and
compare them in \S\ref{sec_discussion}.

\begin{figure}[p]
\newcommand{\tabone  }[1]{\raisebox{0.18cm}{\rule{0cm}{.55cm}#1}}
\newcommand{\tabtwo  }[1]{\raisebox{0.18cm}{\rule{0cm}{.55cm}#1}}
\newcommand{\maxone  }[0]{\tabone{\framebox[0.5cm]{\footnotesize 9}}}
\newcommand{\midone  }[0]{\tabone{\framebox[0.5cm]{\footnotesize 5}}}
\newcommand{\minone  }[0]{\tabone{\framebox[0.5cm]{\footnotesize 0}}}
\newcommand{\opeone  }[0]{\tabone{1}}
\newcommand{\opesix  }[0]{\tabone{6}}
\newcommand{\rowtop  }[2]{#1        &a      &b      &c      &#2         }
\newcommand{\rowone  }[0]{\tabtwo{1}&\minone&\maxone&\opeone&\tabtwo{10}}
\newcommand{\rowtwo  }[0]{\tabtwo{2}&\maxone&\maxone&\minone&\tabtwo{18}}
\newcommand{\rowthree}[0]{\tabtwo{3}&\opesix&\minone&\midone&\tabtwo{11}}
\newcommand{\rowbot  }[1]{#1        &15     &18     &6      &           }
\newcommand{\RowColumnIndex}[0]{\parbox{1.0cm }{\rule[-.10cm]{0cm}{1.00cm}\shortstack{row\\column\\index}}}
\newcommand{\RowSum        }[0]{\hspace{.12cm }\parbox{0.57cm}{\shortstack{row\\sum}}}
\newcommand{\ColumnSum     }[0]{\parbox{1.01cm}{\rule[-.18cm]{0cm}{0.75cm}\shortstack{column\\sum       }}}

            \begin{center}                   
                  \footnotesize
                  \begin{tabular}{|c||c|c|c||c|}                 
                        \hline \rowtop{\RowColumnIndex}{\RowSum} \\
                        \hline \hline \rowone                    \\
                        \hline \rowtwo                           \\
                        \hline \rowthree                         \\
                        \hline \hline \rowbot{\ColumnSum}        \\
                        \hline
                  \end{tabular}
            \end{center}

            \vspace{0.6in}
            \begin{center}
                  \begin{picture}(160,60)(0,0)
                         \setlength{\unitlength}{.1cm}
                         \multiput(12,12)(16,0){3}{\circle*{4}}  
                         \multiput(12,28)(16,0){3}{\circle*{4}}  
                         \put(12,12){\line  ( 1, 0){32}}
                         \put(12,12){\vector(+1, 0){14}}
                         \put(28,12){\vector(+1, 0){14}}
                         \put(12,28){\line  ( 1, 0){32}}
                         \put(44,28){\vector(-1, 0){14}}
                         \put(28,28){\vector(-1, 0){14}}
                         \put(12,12){\line  ( 0, 1){16}}
                         \put(12,28){\vector( 0,-1){14}}
                         \put(28,12){\line  ( 0, 1){16}}
                         \put(28,28){\vector( 0,-1){14}}
                         \put(44,12){\line  ( 0, 1){16}}
                         \put(10, 6){$C_a$}
                         \put(26, 6){$R_2$}
                         \put(42, 6){$C_c$}
                         \put(10,31){$R_1$}
                         \put(26,31){$C_b$}
                         \put(42,31){$R_3$}
                   \end{picture}
            \end{center}

\begin{center}
\parbox{5in}{
The number in a table cell is its value.  A cell with a box is
suppressed.  The lower and upper bounds of the cells are 0 and 9.
The graph is the suppressed graph of the table.  Vertex $R_p$
corresponds to row $p$, and vertex $C_q$ to column $q$.}
\end{center}
\vspace{.2in}\caption{A Table and Its Suppressed Graph.}
\label{figure_suppressed_graph}
\end{figure}

\section{Preliminaries}\label{sec_basics_graph}
Every graph in this paper is a {\it mixed} graph, i.e., it may
contain both undirected and directed edges with at most one edge
between two vertices.  Let $\TT$ be a table.  The {\it suppressed
graph} $\HH=(A,B,E)$ and the {\it total graph} $\HH'=(A,B,E')$ of
$\TT$ are the bipartite graphs constructed here (see
Figure~\ref{figure_suppressed_graph} for an example).  For each
row (respectively, column) of $\TT$, there is a vertex in $A$
(respectively, $B$); this vertex is called a {\it row}
(respectively, {\it column}) vertex.  For each cell $x$ at row
$i$ and column $j$, there is an edge $e \in E'$ between the
vertices of row $i$ and column $j$.  If the value of $x$ is
strictly between its lower and upper bounds, then $e$ is
undirected.  Otherwise, if the value equals the lower
(respectively, upper) bound, then $e$ points from the row
endpoint to the column endpoint (respectively, from the column
endpoint to the row endpoint).  $E$ consists of the edges
corresponding to the suppressed cells.  Note that $\HH$ is a
subgraph of $\HH'$ and $\HH'$ is {\it complete} (i.e., for all $u
\in A$ and $v \in B$, $E'$ has exactly one edge between $u$ and
$v$). Also, given an arbitrary complete bipartite graph and a
subgraph on the same vertices, it takes only linear time to
construct a table with these two graphs as its total and
suppressed graphs.

A {\it traversable} cycle or path is one that can be traversed
along its edge directions.  A {\it direction-blind} cycle or path
is one that can be traversed if its edge directions are
disregarded; we often omit the word direction-blind for brevity.
A graph is {\it connected} if each pair of vertices are in a
path. A {\it connected component} is a maximal connected
subgraph.  A {\it nonsingleton} connected component is one with
two or more vertices.  A graph is {\it strongly connected} if
each pair of vertices are in a traversable cycle.  A {\it strong
component} is a maximal strongly connected subgraph.

The {\it effective area} of a linear invariant $F$ of $\TT$,
denoted by $EA(F)$, is the set of suppressed cells in the nonzero
terms of $F$.  $EA(F)$ is also regarded as a set of edges in
$\HH$.  $F$ is {\it nonzero} if $EA(F) \neq \emptyset$.  $F$ is
{\it minimal} if it is nonzero and $\TT$ has no nonzero linear
invariant whose effective area is a proper subset of $EA(F)$.
Note that given a minimal linear invariant $F$, if $F'$ is a
nonzero linear invariant with $EA(F')
\subseteq EA(F)$, then $F'$ is also minimal and is a multiple of
$F$. Thus, a minimal linear invariant is unique up to a
multiplicative factor with respect to its effective area.

An edge set of a graph is an {\it edge cut} if its removal
disconnects a connected component.  An edge cut is {\it minimal}
if no proper subset of it is an edge cut. 

\begin{fact}\label{fact_char_mec}
Let $Z$ be an edge set of a strong component $\HH'$ of $\HH$.
$Z$ is a minimal edge cut of $\HH'$ if and only if $\HH'-Z$ has
exactly two connected components, say, $\HH_1$ and $\HH_2$, and
each edge of $Z$ is between $\HH_1$ and $\HH_2$.
\end{fact}

Assume that $Z$ is a minimal edge cut of $\HH'$.  $Z$ is {\it
bipartite} if the endpoints of $Z$ in $\HH_1$ are all row
vertices or all column vertices.  An edge set of $\HH$ is a
(respectively, {\it bipartite}) {\it basic set} if it consists of
an edge not in any strong component of $\HH$ or is a
(respectively, bipartite) minimal edge cut of some strong
component.

\begin{theorem}[\cite{Kao92.mli}] \label{thm_mli}
\begin{enumerate}
\item \label{thm_mli_mec} 
A linear invariant of $\TT$ is minimal if and only if its
effective area is a basic set of $\HH$.  Also, for each basic set
$Z$ of $\HH$, there is a minimal linear invariant $F$ of $\TT$
with $EA(F) = Z$.
\item \label{thm_gt_mli_coef}
Every minimal linear invariant is a multiple of a unitary
invariant. Furthermore, a minimal linear invariant $F$ of $\TT$
is a multiple of a sum invariant if and only if $EA(F)$ is a
bipartite basic set of $\HH$.
\item \label{thm_mdthm} 
For each nonzero linear invariant $F$ of $\TT$, there exist
unitary minimal linear invariants $F_1,\ldots,F_k$ of $\TT$ such
that $F=\sum_{i=1}^k c_i{\cdot}F_i$ for some $c_i > 0$,
$EA(F)=\cup_{i=1}^{k}EA(F_i)$, and for each $F_i$ and each $e \in
EA(F_i)$, the coefficients of $e$ in $F$ and $F_i$ are either
both positive or both negative.
\end{enumerate}
\end{theorem}

{\it Remark.} A referee has indicated that a different proof for
Theorem~\ref{thm_mli} from that in~\cite{Kao92.mli} can be
constructed by means of conformal vector decomposition
{\cite{Rockafellar69, Rockafellar84}}.

\section{Protection of a cell}
\label{sec_cell}
A suppressed cell of $\TT$ is {\it protected} if it is not an
invariant cell.

\subsection{Cell protection and bridge-freeness}
\label{subsec_cell_char}
A graph is {\it bridge-free} if it has no edge cut consisting of
a single edge.

\begin{theorem}[\cite{Gusfield88}]\label{thm_cell_char}
\begin{enumerate}
\item
A suppressed cell of $\TT$ is protected if and only if it is an
edge in an edge-simple traversable cycle of $\HH$.
\item
The suppressed cells of $\TT$ are all protected if and only if
each connected component of $\HH$ is strongly connected and
bridge-free.
\end{enumerate}
\end{theorem}

\begin{corollary}[\cite{Gusfield88}]
Given $\HH$, the unprotected cells of $\TT$ can be found in
$O(|\HH|)$ time.
\end{corollary}

\subsection{Optimal suppression problems for cell protection}
\label{subsec_cell_suppression_problem}
The problem below is concerned with suppressing the minimum
number of additional cells in $\TT$ such that the original and
the new suppressed cells in the resulting table are all
protected.

\namedkaoproblem{Protection of All Cells}{\label{problem_cell_table}
\begin{itemize}
\item Input: 
$\TT$ and an integer $p \geq 0$.
\item Output: 
Is there a set $P$ consisting of at most $p$ published cells of
$\TT$ such that all suppressed cells are protected in the table
formed by $\TT$ with the cells in $P$ also suppressed?
\end{itemize}}

Problem~\ref{problem_cell_table} can be reformulated as the graph
augmentation problem below.

\kaoproblem{\label{problem_cell_graph}
\begin{itemize}
\item Input: 
A complete bipartite graph $\HH'$, a subgraph $\HH$, and an
integer $p \geq 0$.
\item Output:
Is there a set $P$ of at most $p$ edges in $\HH'-\HH$ such that
each connected component of $\HH \cup P$ is strongly connected
and bridge-free?
\end{itemize}}

\begin{lemma}\label{lem_problem_cell_table_graph_equiv}
Problems~\ref{problem_cell_table} and \ref{problem_cell_graph}
can be reduced to each other in linear time.
\end{lemma}
\begin{proof}
The proof follows from Theorem~\ref{thm_cell_char}(2).
\end{proof}

The next problem is NP-complete \cite{GJ79}. It is used here to prove that
Problems \ref{problem_cell_table} and
\ref{problem_cell_graph} are hard. 

\namedkaoproblem{Hitting Set}{\label{problem_hit}
\begin{itemize}
\item Input:
A finite set $S$, a nonempty set $W \subseteq 2^S$, and an
integer $h
\geq 0$.
\item Output:
Is there a subset $S'$ of $S$ such that $|S'| \leq h$ and
$S'$ contains at least one element in each set in $W$?
\end{itemize}}

\begin{theorem}\label{thm_cell_npc}
Problems \ref{problem_cell_table} and \ref{problem_cell_graph}
are NP-complete.
\end{theorem}
\begin{proof}
Problems \ref{problem_cell_table} and \ref{problem_cell_graph} are both in
NP.  To prove their completeness, by
Lemma~\ref{lem_problem_cell_table_graph_equiv}, it suffices to reduce
Problem~\ref{problem_hit} to Problem~\ref{problem_cell_graph}.

Given an instance $S=\{s_1,\ldots,s_\alpha\}$,
$W=\{S_1,\ldots,S_\beta\}$, $h$ of Problem~\ref{problem_hit}, an
instance $\HH'=(A,B,E'),\HH=(A,B,E),p$ of
Problem~\ref{problem_cell_graph} is constructed as follows:
\begin{itemize}
\item Rule 1:
Let $A=\{a_0,a_1,\ldots,a_\alpha\}$. The vertices
$a_1,\ldots,a_\alpha$ correspond to $s_1,\ldots,s_\alpha$, but
$a_0$ corresponds to no $s_i$.

\item Rule 2:
Let $B=\{b_0,b_1,\ldots,b_\beta\}$.  The vertices $b_1,\ldots,b_\beta$
correspond to $S_1,\ldots,S_\beta$ of $S$, but $b_0$ corresponds to
no $S_j$.

\item Rule 3:
Let $E'$ consist of the following edges:
\begin{enumerate}
\item 
The edge between $a_0$ and $b_0$ is ${b_0}\rightarrow{a_0}$.

\item 
For all $j$ with $1 \leq j \leq \beta$, the edge between $a_0$ and
$b_j$ is ${a_0}\rightarrow{b_j}$.

\item 
For all $i$ with $1 \leq i \leq \alpha$, the edge between $a_i$ and
$b_0$ is ${a_i}\rightarrow{b_0}$.

\item 
For each $s_i$ and each $S_j$, if $s_i \in S_j$, then the edge
between $a_i$ and $b_j$ is ${b_j}\rightarrow{a_i}$; otherwise it
is ${a_i}\rightarrow{b_j}$.
\end{enumerate}

\item Rule 4:
Let $E=\{{b_0}\rightarrow{a_0}\} \cup \{{a_0}\rightarrow{b_1},
\cdots,{a_0}\rightarrow{b_\beta}\}$.
 
\item  Rule 5:
Let $p=h+\beta$.
\end{itemize}

The above construction can be easily computed in polynomial time.
The next two claims show that it is indeed a desired reduction
from Problem~\ref{problem_hit} to
Problem~\ref{problem_cell_graph}.

\begin{claim}\label{claim_cell_1}
If some $S' \subseteq S$ with $|S'| \leq h$ has at least one
element in each $S_j$, then some $P \subseteq E'-E$ consists of
at most $p$ edges such that every connected component of $\HH
\cup P$ is strongly connected and bridge-free.
\end{claim}

To prove this claim, observe that for each $S_j$, some $s_{i_j} \in S' \cap
S_j$ exists.  By Rule 3(4), $P_1=\{{b_1}\rightarrow{a_{i_1}},\ldots,
{b_\beta}\rightarrow{a_{i_\beta}}\}$ exists.  By Rule 3(3),
$P_2=\{{a_{i_1}}\rightarrow{b_0},\ldots, {a_{i_\beta}}\rightarrow{b_0}\}$
exists.  Let $P=P_1 \cup P_2$.  Note that $P_1$ consists of $\beta$ edges.
$P_2$ consists of at most $|S'|$ edges.  Thus $P$ has at most $p=\beta+h$
edges.  For all $j$ with $1 \leq j \leq
\beta$, the edges ${b_0}\rightarrow{a_0},{a_0}\rightarrow{b_j},
{b_j}\rightarrow{a_{i_j}},{a_{i_j}}\rightarrow{b_0}$ form a vertex-simple
traversable cycle.  Because $E \cup P$ consists of the edges in these
cycles, every connected component of $\HH \cup P$ is strongly connected and
bridge-free. This finishes the proof of Claim~\ref{claim_cell_1}.

\begin{claim}\label{claim_graph_to_hit_cell}
If some $P \subseteq E'-E$ consists of at most $p$ edges such that every
connected component of $\HH \cup P$ is strongly connected and bridge-free,
then some $S' \subseteq S$ with $|S'|
\leq h$ has at least one element in each $S_j$.
\end{claim}

To prove this claim, observe that for all $j$ with $1 \leq j \leq \beta$,
by Rule 4, $E$ contains ${a_0}\rightarrow{b_j}$ but no edge pointing from
$b_j$.  Because every connected component of $\HH \cup P$ is strongly
connected, $P$ contains an edge ${b_j}\rightarrow{a_{i_j}}$ for some $i_j$.
By Rule 3(4), $s_{i_j} \in S_j$.  Let $S'=\{s_{i_1},\ldots,s_{i_\beta}\}$.
Note that $P$ contains
${b_1}\rightarrow{a_{i_1}},\ldots,{b_\beta}\rightarrow{a_{i_\beta}}$ but
$E$ contains no edges pointing from $\{a_{i_1},\ldots,a_{i_\beta}\}$.
Because every connected component of $\HH \cup P$ is strongly connected,
$P$ must also contain at least one edge pointing from each vertex in
$\{a_{i_1},\ldots,a_{i_\beta}\}$.  Thus $P$ contains at least $|S'|+\beta$
edges.  Then $|S'| \leq h$ because $|P|\leq\beta+h$. This finishes the
proof of Claim~\ref{claim_graph_to_hit_cell} and thus that of
Theorem~\ref{thm_cell_npc}.
\end{proof}

The next two problems are optimization versions of
Problems~\ref{problem_cell_table} and \ref{problem_cell_graph}
for undirected graphs and tables whose total graphs are
undirected.

\namedkaoproblem{Protection of All
Cells}{\label{problem_cell_undirected_table}
\begin{itemize}
\item Input:
The suppressed graph of a table $\TT$ whose total graph is
undirected.
\item Output:
A set $P$ consisting of the smallest number of published cells of
$\TT$ such that all suppressed cells are protected in the table
formed by $\TT$ with the cells in $P$ also suppressed.
\end{itemize}}

\kaoproblem{\label{problem_cell_undirected_graph}
\begin{itemize}
\item Input:
A bipartite undirected graph $\HH=(A,B,E)$.
\item Output:
A set $P$ consisting of the smallest number of undirected edges
between $A$ and $B$ but not in $E$ such that every connected
component of $(A,B,E \cup P)$ is bridge-free.
\end{itemize}}

Note that Problem~\ref{problem_cell_undirected_graph} needs not
specify $\HH'$ because it is undirected and thus is unique for
$\HH$.  Similarly, $\HH \cup P$ is always strongly connected.

\begin{lemma}
\label{lem_problem_cell_undirected_table_graph_equiv}
Problems~\ref{problem_cell_undirected_table} and
\ref{problem_cell_undirected_graph} can be reduced to each other
in linear time.
\end{lemma}
\begin{proof}
The proof is similar to that of
Lemma~\ref{lem_problem_cell_table_graph_equiv}.
\end{proof}

\begin{theorem}[\cite{Gusfield87}]
Problem {\ref{problem_cell_undirected_graph}} is solvable in
linear time; thus so is
Problem~\ref{problem_cell_undirected_table}.
\end{theorem}

\section{Protection of rows and columns}
This section discusses the data security of a table at Levels 2
and 3 in a unified framework.  Let $EA(R)$ denote the set of
suppressed cells in a row or column $R$.  Let $\MyGamma{R} =
\sum_{e \in EA(R)} e$.  Let $R_1,\ldots,R_k$ be $k$ rows or $k$
columns of $\TT$, but no mixed case.  For Level 3 data security,
$\{R_1,\ldots,R_k\}$ is {\it protected} with respect to the
linear invariants (respectively, the positive invariants, the
unitary invariants, the sum invariants, or the rectangular sum
invariants) if the conditions below hold:
\begin{enumerate}
\item 
Each linear invariant (respectively, positive invariant, unitary
invariant, sum invariant, or rectangular sum invariant) $F$ of
$\TT$ with $EA(F) \subseteq
\cup_{i=1}^k EA(R_i)$ is a linear combination of
$\MyGamma{R}_1,\ldots,\MyGamma{R}_k$.
\item 
No suppressed cell of $R_1,\ldots,R _k$ is an invariant cell.
\end{enumerate}
Level 2 data security is a special case of Level 3 with $k=1$ and
its definitions can be simplified.  A row or column $R$ is {\it
protected} with respect to the linear invariants (respectively,
the positive invariants, the unitary invariants, or the sum
invariants) if the conditions below hold:
\begin{enumerate}
\item 
Each linear invariant (respectively, positive invariant, unitary
invariant, or sum invariant) $F$ with $EA(F) \subseteq EA(R)$ is
a multiple of~$\MyGamma{R}$.
\item 
No suppressed cell in $R$ is an invariant cell.
\end{enumerate}
We do not explicitly consider the protection of $R$ with respect
to the rectangular sum invariants because for $k=1$ these
invariants are the same as the sum invariants.  Also, the five
types of invariants here are implicitly considered for cell
protection because a linear invariant with exactly one nonzero
term is essentially an invariant cell.

The two conditions in the definitions are based on technical
considerations.  No matter how many cells in $\TT$ are
suppressed, $\MyGamma{R}_1,\ldots,\MyGamma{R}_k$ and their linear
combinations are always linear invariants.  Thus the first
condition gives the best possible protection for $R_1,\ldots,R_k$
as a whole.  If $R_i$ has either no suppressed cell or at least
two, the first condition implies the second one; otherwise, the
first condition holds trivially but the only suppressed cell in
$R_i$ is an invariant.  The second condition is adopted to avoid
this undesirable situation.

These definitions also require that $R_1,\ldots,R_k$ be all rows
or all columns.  In these two pure cases,
$EA(R_1),\ldots,EA(R_k)$ are pairwise disjoint.  Therefore, a
linear combination of $\MyGamma{R}_1,\ldots,\MyGamma{R}_k$ has a
very simple structure and encodes essentially the same
information as do $\MyGamma{R}_1,\ldots,\MyGamma{R}_k$.  In
contrast, if at least one $R_i$ is a row and at least one $R_j$
is a column, then a linear combination of
$\MyGamma{R}_1,\ldots,\MyGamma{R}_k$ may have a very complex
structure and may encode very different information from that
contained in $\MyGamma{R}_1,\ldots,\MyGamma{R}_k$.  Furthermore,
unlike in the two pure cases, these definitions do not seem to
have useful characterizations in the mixed case.

The importance of the first four types of invariants considered
in the definitions are evident.  The fifth type, a rectangular
sum invariant, is motivated by a popular technique for protecting
information in a table. Let $e$ be an invariant cell at row $i$
and column $j$.  To protect $e$, row $i$ can be split into
several rows, and column $j$ into several columns.
Correspondingly, $e$ is split into four or more cells.  Then
enough of these refined cells can be suppressed to ensure that
each suppressed refined cell is protected.  However, the sum of
the suppressed refined cells of $e$ is a rectangular sum
invariant.  This property can be used to uniquely determine the
value of $e$.  Thus the consideration of rectangular sum
invariants renders this refinement approach useless at the third
level of data security.

\subsection{Equivalence of $k$ row-column protection}
\label{subsec_k_row_column_equiv} 
This section shows that the five definitions of $k$ row-column
protection are all equivalent.

\begin{lemma}\label{lem_sum_minimal_rectangular}
Every sum minimal invariant is rectangular.
\end{lemma}
\begin{proof}
Let $F$ be a sum minimal invariant of $\TT$. If $EA(F)$ consists
of an edge not in any strong component of $\HH$, then $F$ is
trivially rectangular.  Otherwise, by Theorem~\ref{thm_mli}
$EA(F)$ is a bipartite minimal cut set of a strong component
$\HH'$ of $\HH$.  By Fact~\ref{fact_char_mec}, $\HH'-EA(F)$ has
two connected components $\HH'_1$ and $\HH'_2$.  Let $U_1$ and
$U_2$ be the sets of endpoints of $EA(F)$ in $\HH'_1$ and
$\HH'_2$, respectively.  By the bipartiteness of $EA(F)$, without
loss of generality the vertices in $U_1$ are rows in $\TT$ and
those in $U_2$ are columns.  Then $F$ is rectangular because
$EA(F)$ consists of the edges between $U_1$ and $U_2$ in $\HH$.
\end{proof}

\begin{lemma}\label{lem_k_row_column_equiv_positive}
If no $EA(R_i)$ is empty, the statements below are equivalent:
\begin{enumerate}
\item 
Every positive invariant $F$ with $EA(F) \subseteq
\cup_{i=1}^k EA(R_i)$ is a linear combination of
$\MyGamma{R}_1,\ldots,\MyGamma{R}_k$.
\item 
Every sum invariant $F$ with $EA(F) \subseteq
\cup_{i=1}^k EA(R_i)$ is a linear combination of
$\MyGamma{R}_1,\ldots,\MyGamma{R}_k$.
\item 
Every rectangular sum invariant $F$ with $EA(F) \subseteq
\cup_{i=1}^k EA(R_i)$ is a linear combination of
$\MyGamma{R}_1,\ldots,\MyGamma{R}_k$.
\item 
$\MyGamma{R}_1,\ldots,\MyGamma{R}_k$ are the only sum minimal
invariants of $\TT$ whose effective areas are subsets of
$\cup_{i=1}^k EA(R_i)$.
\end{enumerate}
\end{lemma}
\begin{proof}
The directions $1 \Rightarrow 2 \Rightarrow 3$ are
straightforward.  The direction $4 \Rightarrow 1$ follows from
the fact that by Statement 4 $\MyGamma{R}_1,\ldots,\MyGamma{R}_k$
are the only factors in the decomposition in
Theorem~\ref{thm_mli}(3) for a positive invariant $F$ with $EA(F)
\subseteq \cup_{i=1}^k EA(R_i)$.  To prove $3 \Rightarrow 4$,
note that because $\MyGamma{R}_j$ is a positive invariant for all
$R_j$, by Theorem~\ref{thm_mli}(3) there is a sum minimal
invariant $F$ with $EA(F) \subseteq EA(\MyGamma{R}_j)$. Since $F$
is also rectangular, by Statement 3, $F=\sum_{i=1}^k
c_i{\cdot}\MyGamma{R}_i$ for some $c_i$.  Because
$\MyGamma{R}_1,\ldots,\MyGamma{R}_k$ share no variable, by the
minimality of $F$ and coefficient comparison $\MyGamma{R}_j$
equals $F$ and thus is a sum minimal invariant.  To prove the
desired uniqueness of $\MyGamma{R}_1,\ldots,\MyGamma{R}_k$, let
$F'$ be a sum minimal invariant with $EA(F') \subseteq
\cup_{i=1}^k EA(R_i)$.  By
Lemma~{\ref{lem_sum_minimal_rectangular}}, $F'$ is rectangular.
By Statement 3, $F'=\sum_{i=1}^k c'_i{\cdot}\MyGamma{R}_i$ for
some $c'_i$.  Because $F'$ is nonzero, some $c'_h \neq 0$.
Because $\MyGamma{R}_1,\ldots,\MyGamma{R}_k$ do not share
variables, $EA(\MyGamma{R}_h) \subseteq EA(F')$.  Then,
$F'=\MyGamma{R}_h$ by coefficient comparison and the minimality
of $F'$.
\end{proof}

\begin{lemma}\label{lem_k_row_column_equiv_general}
If no $EA(R_i)$ is empty, the statements below are equivalent:
\begin{enumerate}
\item 
Every linear invariant  of $\TT$ whose effective area is a subset of
$\cup_{i=1}^k EA(R_i)$ is a linear combination of
$\MyGamma{R}_1,\ldots,\MyGamma{R}_k$.
\item 
Every unitary invariant whose effective area is a subset of
$\cup_{i=1}^k EA(R_i)$ is a linear combination of
$\MyGamma{R}_1,\ldots,\MyGamma{R}_k$.
\item 
$\MyGamma{R}_1,\ldots,\MyGamma{R}_k$ and their nonzero multiples
are the only minimal linear invariants of $\TT$ whose effective
areas are subsets of $\cup_{i=1}^k EA(R_i)$.
\end{enumerate}
\end{lemma}
\begin{proof}
The proof is similar to that of
Lemma~\ref{lem_k_row_column_equiv_positive}.
\end{proof}

\begin{lemma}\label{lem_k_row_column_char_1}
$\{R_1,\ldots,R_k\}$ is protected with respect to the positive
invariants $($respectively, the linear invariants$)$ if and only
the following statements hold:
\begin{enumerate}
\item
For each strong component $D$ of $\HH$ and each vertex $R_i$
contained in $D$, the component $D$ contains all edges incident
to $R_i$~in~$\HH$.
\item
The nonempty sets among $EA(R_1),\ldots,EA(R_k)$ are the only
bipartite minimal edge cuts $($respectively, the only minimal
edge cuts$)$ of the strong components of $\HH$ among the subsets
of $\cup_{i=1}^k EA(R_i)$.
\item
Each vertex $R_i$ is either isolated or incident to two or more
edges~in~$\HH$.
\end{enumerate}
\end{lemma}
\begin{proof}
The proof of the lemma for the positive invariants and that for
the general invariants are similar; only the former is detailed
here.  For the direction $\Rightarrow$, Statement 3 follows from
the second condition of the definition of $\{R_1,\ldots,R_k\}$
being protected.  Then, Statements 1 and 2 follows from
Lemma~\ref{lem_k_row_column_equiv_positive}(1),
\ref{lem_k_row_column_equiv_positive}(4), and 
Theorem~\ref{thm_mli}(1), \ref{thm_mli}(2). For the direction
$\Leftarrow$, by Statements 1 and 2, Theorem~\ref{thm_mli} and
Lemma~\ref{lem_k_row_column_equiv_positive}, the first condition
of $\{R_1,\ldots,R_k\}$ being protected is satisfied.  The second
condition then follows from Statement 3.
\end{proof}

A set of vertices in a connected graph is a {\it vertex cut} if
its removal disconnects the graph.

\begin{fact}\label{fact_k_row_column_char_equiv_graph}
If each $EA(R_i)$ is included in the strong component of $\HH$
that contains $R_i$, then the following statements are
equivalent:
\begin{enumerate}
\item
Among the subsets of $\cup_{i=1}^k EA(R_i)$, the nonempty sets
$EA(R_i)$ are the only minimal edge cuts of the strong components
of $\HH$.
\item
Among the subsets of $\cup_{i=1}^k EA(R_i)$, the nonempty sets
$EA(R_i)$ are the only bipartite minimal edge cuts of the strong
components of $\HH$.
\item 
$\{R_1,\ldots,R_k\}$ includes no vertex cut of any strong
component~of~$\HH$.
\end{enumerate}
\end{fact}

\begin{theorem}\label{thm_k_row_column_equiv}
The five definitions of a set of $k$ rows or $k$ columns being
protected are all equivalent.
\end{theorem}
\begin{proof}
If some $EA(R_i)=\emptyset$, then $\{R_1,\ldots,R_k\}$ is
protected if and only if $\{R_1,\ldots,R_k\}-\{R_i\}$ is
protected.  Thus without loss of generality assume that no
$EA(R_i)$ is empty.  Then, by
Lemma~\ref{lem_k_row_column_equiv_positive} the protection
definitions with respect to the positive, sum, and rectangular
invariants are all equivalent.  Similarly, by
Lemma~\ref{lem_k_row_column_equiv_general}, those with respect to
the general and unitary invariants are also equivalent.  This
theorem then follows directly from
Lemma~\ref{lem_k_row_column_char_1} and
Fact~\ref{fact_k_row_column_char_equiv_graph}.
\end{proof}

\subsection{$k$ Row-column protection and
bipartite-$(k+1)$-connectivity}
\label{subsec_k_row_column_char}
A connected bipartite graph $\GG = (X,Y,I)$ is {\it
bipartite-$(k+1)$-connected} if $|X| \geq k+1$, $|Y| \geq k+1$,
and neither $X$ nor $Y$ includes a vertex cut of at most $k$
vertices. $\GG$ is {\it $(k+1)$-connected} if $|X \cup Y| \geq
k+1$ and there is no vertex cut of at most $k$ vertices.

\begin{lemma}\label{lem_k_row_column_char_2}
$\{R_1,\ldots,R_k\}$ is protected if and only if the statements
below hold:
\begin{enumerate}
\item 
For each strong component $D$ of $\HH$ and each vertex $R_i \in
D$, $D$ contains all the edges incident to $R_i$ in $\HH$.
\item
$\{R_1,\ldots,R_k\}$ includes no vertex cut of any strong
component~of~$\HH$.
\item 
Each $R_i$ is either isolated or incident to two or more edges in
$\HH$.
\end{enumerate}
\end{lemma}
\begin{proof}
This lemma follows from Theorem~\ref{thm_k_row_column_equiv},
Lemma~\ref{lem_k_row_column_char_1} and
Fact~\ref{fact_k_row_column_char_equiv_graph}.
\end{proof}

\begin{theorem}\label{thm_k_row_column_char_all}
Every set of at most $k$ rows or $k$ columns of $\TT$ is
protected if and only if every nonsingleton connected component
of $\HH$ is strongly connected and bipartite-$(k+1)$-connected.
\end{theorem}
\begin{proof}
This theorem follows directly from
Lemma~\ref{lem_k_row_column_char_2}.
\end{proof}

\begin{corollary}
\begin{enumerate}
\item
Given $\HH$ and $\{R_1,\ldots,R_k\}$, whether
$\{R_1,\ldots,R_k\}$ is protected can be determined in
$O(|\HH|)$ time.
\item
Give $\HH$ and $k$, whether $\TT$ has any unprotected set of at most $k$
rows or $k$ columns can be answered in $O(k^4n^2)$ time, where $n$ is the
number vertices in $\HH$.
\end{enumerate}
\end{corollary}
\begin{proof}
Statement 1 follows from Lemma~\ref{lem_k_row_column_char_2} in a
straightforward manner using linear-time algorithms for
connectivity and strong connectivity \cite{CLR91}.  Statement 2
follows from Theorem~\ref{thm_k_row_column_char_all}.  The key
step is to test the bipartite-$(k+1)$-connectivity of $\HH$
within the stated time bound.  We first construct two auxiliary
graphs $\HH_A$ and $\HH_B$.  For each vertex $u \in A$, replace
$u$ with $k+1$ copies in $\HH_A$.  For each $u \in A$ and each
edge $e$ in $\HH$ between $u$ and a vertex $v \in B$, replace $e$
with $k+1$ copies between $v$ and the $k+1$ copies of $u$ in
$\HH_A$.  $\HH_B$ is obtained by exchanging $A$ and $B$ in the
construction.  Because $\HH$ is connected and each vertex in $A$
is duplicated $k+1$ times, $\HH_A$ has a vertex cut $U$ of at
most $k$ vertices if and only if $U$ is a subset of $B$ and is a
vertex cut of $\HH$.  A symmetrical statement for $B$ also holds.
Thus $\HH$ is bipartite-$(k+1)$-connected if and only if both
$\HH_A$ and $\HH_B$ are $(k+1)$-connected.  This corollary then
follows from the fact \cite{CKaoT93, NI92} that the
$(k+1)$-connectivity of an $m$-vertex graph can tested in
$O(k^2m^2)$ time if $k \leq \sqrt{m}$.
\end{proof}

\begin{corollary}
Given $\HH$, it takes $O(|\HH|)$ time to find the unprotected
rows and columns of $\TT$ and decide whether all individual rows
and columns of $\TT$ are protected.
\end{corollary}
\begin{proof} 
This corollary follows from Lemma~\ref{lem_k_row_column_char_2} in a
straightforward manner using linear-time algorithms for strong connectivity
and 2-connectivity {\cite{CLR91}}.
\end{proof}

\subsection{Optimal suppression problems for $k$ row-column
protection}\
\label{subsec_k_row_column_suppression_problem}

\namedkaoproblem{Protection of All
Sets}{\label{problem_k_row_column_table}
\begin{itemize}
\item Input: 
$\TT$ and two integers $k>0$ and $p \geq 0$.
\item Output: 
Is there a set $P$ consisting of at most $p$ published cells of
$\TT$ such that every set of at most $k$ rows or $k$ columns is
protected in the table formed by $\TT$ with the cells in $P$ also
suppressed?
\end{itemize}}

Problem~\ref{problem_k_row_column_table} can be reformulated as
the following graph augmentation problem.

\kaoproblem{\label{problem_k_row_column_graph}
\begin{itemize}
\item Input: 
A complete bipartite graph $\HH'$, a subgraph $\HH$, and integers
$k >0$ and $p \geq 0$.
\item Output:
Is there a set $P$ of at most $p$ edges in $\HH'-\HH$ such that
each nonsingleton connected component of $\HH \cup P$ is strongly
connected and bipartite-$(k+1)$-connected?
\end{itemize}}

\begin{lemma}\label{lem_problem_k_row_column_table_graph_equiv}
Problems~\ref{problem_k_row_column_table} and
\ref{problem_k_row_column_graph} can be reduced to each other in
linear time.
\end{lemma}
\begin{proof}
The proof follows from Theorem~\ref{thm_k_row_column_char_all}.
\end{proof}

\begin{theorem}\label{thm_k_row_column_npc}
For $k =1$, Problems \ref{problem_k_row_column_table} and
\ref{problem_k_row_column_graph} are NP-complete. Thus, both
problems are NP-complete for general $k$.
\end{theorem}
\begin{proof}
Problems \ref{problem_k_row_column_table} and
\ref{problem_k_row_column_graph} are both in NP. To prove their
completeness for $k=1$, by
Lemma~\ref{lem_problem_k_row_column_table_graph_equiv}, it
suffices to reduce Problem~\ref{problem_hit} to
Problem~\ref{problem_k_row_column_graph} with $k=1$.  Given an
instance $S=\{s_1,\ldots,s_\alpha\}$, $W=\{S_1,\ldots,S_\beta\}$,
$h$ of Problem~\ref{problem_hit}, let
$\HH'=(A,B,E'),\HH=(A,B,E),p$ be the instance constructed for
Theorem~\ref{thm_cell_npc}.  The next two claims show that this
transformation is indeed a desired reduction.

\begin{claim}\label{claim_row_column_npc_1}
If some $S' \subseteq S$ with $|S'| \leq h$ has at least one
element in each $S_j$, then some $P \subseteq E'-E$ consists of
at most $p$ edges such that every nonsingleton connected
component of $\HH \cup P$ is strongly connected and
bipartite-2-connected.
\end{claim}

To prove this claim, observe that for each $S_j$, some $s_{i_j} \in S' \cap
S_j$ exists.  Let $P_1=\{{b_1}\rightarrow{a_{i_1}},\ldots,
{b_\beta}\rightarrow{a_{i_\beta}}\}$, which exists by Rule 3(4) of the
construction of $\HH'$, $\HH$, and $p$.  By Rule 3(3),
$P_2=\{{a_{i_1}}\rightarrow{b_0},\ldots, {a_{i_\beta}}\rightarrow{b_0}\}$
exists.  Let $P=P_1 \cup P_2$.  Note that $P_1$ consists of $\beta$ edges.
$P_2$ consists of at most $|S'|$ edges.  Thus $P$ has at most $p=\beta+h$
edges. For all $j$ with $1 \leq j \leq \beta$, the edges
${b_0}\rightarrow{a_0},{a_0}\rightarrow{b_j},
{b_j}\rightarrow{a_{i_j}},{a_{i_j}}\rightarrow{b_0}$ form a vertex-simple
traversable cycles.  These cycles all go through $b_0 \rightarrow a_0$ and
form the only nonsingleton connected component of $\HH \cup P$. This
component is clearly strongly connected and bipartite-2-connected.  This
finishes the proof of Claim~\ref{claim_row_column_npc_1}.

\begin{claim}\label{claim_row_column_npc_2}
If some $P \subseteq E'-E$ consists of at most $p$ edges such
that every nonsingleton connected component of $\HH \cup P$ is
strongly connected and bipartite-2-connected, then some $S'
\subseteq S$ with $|S'|$ $\leq h$ has at least one element in
each $S_j$.
\end{claim}

The proof of this claim is the same as that of
Claim~\ref{claim_graph_to_hit_cell}, and uses only the componentwise strong
connectivity of $\HH \cup P$. This finishes the proof of
Theorem~\ref{thm_k_row_column_npc}.
\end{proof}

The next two problems are variants of
Problems~\ref{problem_k_row_column_table} and
{\ref{problem_k_row_column_graph}}.

\namedkaoproblem{Protection of All Sets}{
\label{problem_k_row_column_undirected_table} 
\begin{itemize}
\item Input:
The suppressed graph of a table $\TT$ whose total graph is
undirected, and a positive integer $k$.
\item Output:
A set $P$ consisting of the smallest number of published cells of
$\TT$ such that every set of at most $k$ rows or $k$ columns is
protected in the table formed by $\TT$ with the cells in $P$ also
suppressed.
\end{itemize}}

\kaoproblem{\label{problem_k_row_column_undirected_graph}
\begin{itemize}
\item Input:
A bipartite undirected graph $\HH =(A,B,E)$ and a positive
integer $k$.
\item Output:
A set $P$ consisting of the smallest number of undirected edges
between $A$ and $B$ but not in $E$ such that every nonsingleton
connected component of $(A,B,E \cup P)$ is
bipartite-$(k+1)$-connected.
\end{itemize}}

\begin{lemma}
\label{lem_problem_k_row_column_undirected_table_graph_equiv}
Problems~\ref{problem_k_row_column_undirected_table} and
{\ref{problem_k_row_column_undirected_graph}} can be reduced to
each other in linear time.
\end{lemma}
\begin{proof}
The proof is similar to that of
Lemma~\ref{lem_problem_k_row_column_table_graph_equiv}.
\end{proof}

\begin{theorem}[\cite{Kao93.augbi}]
\label{thm_k_row_column_undirected_graph} 
For $k=1$, Problem \ref{problem_k_row_column_undirected_graph}
can be solved in linear time.
\end{theorem}

\begin{theorem}
\label{thm_k_row_column_undirected_table} 
For $k=1$, Problem \ref{problem_k_row_column_undirected_table}
can be solved in linear time.
\end{theorem}
\begin{proof}
The proof follows from
Lemma~\ref{lem_problem_k_row_column_undirected_table_graph_equiv}
and Theorem~\ref{thm_k_row_column_undirected_graph}.
\end{proof}

\section{Protection of a table}\label{sec_table}
Let $R_1,\ldots,R_n$ be the rows and columns of $\TT$.  $\TT$ is
{\it protected} with respect to the positive invariants
(respectively, the sum invariants, or the rectangular sum
invariants) if it holds the conditions below:
\begin{enumerate}
\item 
Every positive invariant (respectively, nonzero sum invariant, or
nonzero rectangular sum invariant) of $\TT$ is a positive linear
combination of $\MyGamma{R}_1,\ldots,\MyGamma{R}_n$, where a {\it
positive} linear combination is one that has no negative
coefficients and at least one positive coefficient.
\item 
$\TT$ has no invariant cell.
\end{enumerate}
These definitions allow only positive linear combinations,
because general linear combinations of
$\MyGamma{R}_1,\ldots,\MyGamma{R}_n$ generate all linear
invariants and leave nothing for protection. This restriction
excludes the protection with respect to the general linear
invariants.  As a result, the protection with respect to the
unitary invariants are also not considered, because by
Theorem~\ref{thm_mli}, these invariants have the same structures
as the general linear invariants do.

\begin{theorem}\label{thm_table_equiv}
The three definitions of a table being protected are all
equivalent.
\end{theorem}
\begin{proof}
Because a protected table has no invariant cells, each row or
column has either no suppressed cell or at least two suppressed
cells.  It suffices to prove that if $\TT$ holds this condition,
then the statements below are equivalent:
\begin{enumerate}
\item 
Every positive invariant is a positive linear combination of
$\MyGamma{R}_1,\ldots,\MyGamma{R}_n$.
\item 
Every nonzero sum invariant is a positive linear combination of
$\MyGamma{R}_1,\ldots,\MyGamma{R}_n$.
\item 
Every nonzero rectangular sum invariant of $\TT$ is a positive
linear combination of $\MyGamma{R}_1,\ldots,\MyGamma{R}_n$.
\item 
The nonzero linear invariants among
$\MyGamma{R}_1,\ldots,\MyGamma{R}_n$ are the only sum minimal
invariants of $\TT$.
\end{enumerate}

The directions $1 \Rightarrow 2$ and $2 \Rightarrow 3$ are
straightforward. The direction $4 \Rightarrow 1$ follows from
Theorem~\ref{thm_mli}(3). To prove the direction $3 \Rightarrow
4$, note that for each $R_j$ with $EA(R_j)\neq\emptyset$,
$\MyGamma{R}_j$ is a nonzero sum invariant.~By
Theorem~\ref{thm_mli} there is a sum minimal invariant $F$ with
$EA(F) \subseteq EA(\MyGamma{R}_j)$.  $F$ is also rectangular.
By Statement 3, $F=\sum_{i=1}^k c_i{\cdot}\MyGamma{R}_i$ where
$c_i \geq 0$. By coefficient comparison there is some $c_h > 0$
with $EA(R_h)\neq\emptyset$.  Because $c_i \geq 0$, $\emptyset
\neq EA(\MyGamma{R}_h) \subseteq EA(F) \subseteq
EA(\MyGamma{R}_j)$.  Then $R_h=R_j$ because two distinct $R_i$
cannot share more than one cell and each nonempty $EA(R_i)$
contains at least two cells.  Thus $\MyGamma{R}_j$ equals $F$ and
is a sum minimal invariant.  To prove the desired uniqueness of
$\MyGamma{R}_1,\ldots,\MyGamma{R}_n$, let $F'$ be a sum minimal
invariant with $EA(F')\subseteq\cup_{i=1}^k EA(R_i)$.  By
Lemma~\ref{lem_sum_minimal_rectangular}, $F'$ is rectangular.  By
Statement 3, $F'=\sum_{i=1}^k c'_i{\cdot}\MyGamma{R}_i$ where
$c'_i \geq 0$.  By coefficient comparison there is some $c'_j >
0$ with $EA(\MyGamma{R}_j)\neq\emptyset$.  Because $c'_i \geq 0$,
$EA(\MyGamma{R}_j) \subseteq EA(F')$.  Then $F'=\MyGamma{R}_j$ by
coefficient comparison and the minimality of $F'$.
\end{proof}

\subsection{Table protection and bipartite-completeness}
\label{subsec_table_char}
A graph $\GG = (X, Y ,I)$ is {\it bipartite-complete} if it is
complete, $|X| \geq 2$ and $|Y| \geq 2$.

\begin{fact}\label{fact_complete_versus_mec}
Let $u_1,\ldots,u_g$ be the vertices in $\GG$. Let $EA(u_i)$ be
the set of edges incident to $u_i$.  Then $\GG$ is
bipartite-complete if and only if it is bridge-free and has more
than one vertex, and the sets $EA(u_i)$ are its only bipartite
minimal edge cuts.
\end{fact}

\begin{theorem}\label{thm_table_char}
$\TT$ is protected if and only if each nonsingleton connected
component of $\HH$ is strongly connected and bipartite-complete.
\end{theorem}
\begin{proof}
By Fact~\ref{fact_complete_versus_mec}, it suffices to prove that
the following statments are equivalent:
\begin{enumerate}
\item 
$\TT$ is protected.
\item 
The nonzero invariants among $\MyGamma{R}_1,\ldots,\MyGamma{R}_n$ are the
only sum minimal invariants of $\TT$. Also each $R_i$ contains either no
suppressed cell or at least two suppressed cells.
\item 
Each connected component of $\HH$ is strongly connected and
bridge-free. Also the nonempty sets among $EA(R_1),\ldots,EA(R_n)$ are the
only bipartite minimal edge cuts of the strong components of $\HH$.
\end{enumerate}
The equivalence $1 \Leftrightarrow 2$ follows from the proof of
Theorem~\ref{thm_table_equiv}.  The equivalence $2
\Leftrightarrow 3$ follows from Theorems~\ref{thm_mli} and
{\ref{thm_cell_char}}.
\end{proof}

\begin{corollary}
Given $\HH$, it takes linear time in the size of $\HH$ to
determine whether $\TT$ is protected.
\end{corollary}
\begin{proof}
This is an immediate corollary of Theorem~\ref{thm_table_char}.
\end{proof}

\subsection{Optimal suppression problems for table protection}\
\label{subsec_table_suppression_problem}

\namedkaoproblem{Protection of a Table}{\label{problem_table_table}
\begin{itemize}
\item Input: 
$\TT$ and a nonnegative integer $p$.
\item Output: 
Is there a set $P$ consisting of at most $p$ published cells of
$\TT$ such that the table formed by $\TT$ with the cells in $P$
also suppressed is protected?
\end{itemize}}

Problem~\ref{problem_table_table} can be reformulated as the
following graph augmentation problem.

\kaoproblem{\label{problem_table_graph}
\begin{itemize}
\item Input: 
A complete bipartite graph $\HH'$, a subgraph $\HH$, and an
integer $p \geq 0$.
\item Output: 
Is there a set $P$ of at most $p$ edges in $\HH'-\HH$ such that
each nonsingleton connected component $\HH
\cup P$ is strongly connected and
bipartite-complete?
\end{itemize}}

\begin{lemma}\label{lem_problem_table_table_graph_equiv}
Problems~\ref{problem_table_table} and
\ref{problem_table_graph} can be reduced to each other in
linear time.
\end{lemma}
\begin{proof}
The proof follows from Theorem~\ref{thm_table_char}.
\end{proof}

\begin{theorem}\label{thm_table_npc}
Problems~\ref{problem_table_table} and
{\ref{problem_table_graph}} are NP-complete.
\end{theorem}
\begin{proof}
Problems \ref{problem_table_table} and {\ref{problem_table_graph}} are both
in NP. To prove their completeness, by
Lemma~\ref{lem_problem_table_table_graph_equiv}, it suffices to reduce
Problem~\ref{problem_hit} to Problem~\ref{problem_table_graph}.  Given an
instance $S=\{s_1,\ldots,s_\alpha\}$, $W=\{S_1,\ldots,S_\beta\}$, $h$ of
Problem~\ref{problem_hit}, let $\HH'=(A,B,E'),\HH=(A,B,E),p$ be the
instance constructed for Theorem~\ref{thm_cell_npc} with the modification
below:
\begin{itemize}
\item  Rule $5'$: Let $p=(\beta+1){\cdot}h$.
\end{itemize}

This construction can be computed in polynomial time.  The next
two claims show that it is a desired reduction from
Problem~\ref{problem_hit} to Problem~\ref{problem_table_graph}.

\begin{claim}\label{claim_thm_table_1}
If some $S' \subseteq S$ with $|S'| \leq h$ has at least one element in
each $S_j$, then some $P \subseteq E'-E$ consists of at most $p$ edges such
that every nonsingleton connected component of $\HH \cup P$ is strongly
connected and bipartite-complete.
\end{claim}

To prove this claim, observe that for each $S_j$, some $s_{i_j} \in S' \cap
S_j$ exists.  Let $A'=\{a_{i_1},\ldots,a_{i_\beta}\}$.  Let
$B'=\{b_1,\ldots,b_\beta\}$.  Let $P_1$ be the set of edges in $E'$ from
$B'$ to $A'$.  Let $P_2$ be the set of edges in $E'$ from $A'$ to $b_0$.
Let $P = P_1 \cup P_2$.  Note that $P$ has at most $p=(\beta+1){\cdot}h$
edges because $A'$ has at most $|S'| \leq h$ vertices.  For each $j$ with
$1 \leq j \leq \beta$, the edge $b_j \rightarrow a_{i_j}$ is in $P_1$ by
Rule 3(4) of the construction of $\HH'$, $\HH$, and $p$.  Also, $b_0
\rightarrow a_0$, ${a_0}\rightarrow{b_j}$, and
${a_{i_j}}\rightarrow{b_0}$ are in $\HH \cup P$.  These four edges form a
vertex-simple traversable cycle.  These cycles form the only nonsingleton
connected component in $\HH \cup P$.  Because these cycles all go through
$a_0$, this component is strongly connected.  By the choice of $P$, this
component is bipartite-complete. This finishes the proof of
Claim~\ref{claim_thm_table_1}.

\begin{claim}\label{claim_thm_table_2}
If some $P \subseteq E'-E$ consists of at most $p$ edges such that every
nonsingleton connected component of $\HH \cup P$ is strongly connected and
bipartite-complete, then some $S' \subseteq S$ with $|S'| \leq h$ has at
least one element in each $S_j$.
\end{claim}

To prove this claim, observe that because every connected component of $\HH
\cup P$ is strongly connected, for each $j$ with $1 \leq j \leq \beta$, the
set $P$ contains some edges $b_j \rightarrow a_{i_j}$ and $a_{i_j}
\rightarrow b_{j'}$.  Then $i_j \neq 0$ and $s_{i_j}$ exists in
$S_j$ by Rule 3 of the construction of $\HH'$, $\HH$, and $p$.  Let
$S'=\{s_{i_1},\ldots,s_{i_\beta}\}$.  Let $D$ be the connected component of
$\HH \cup P$ that contains $a_0$. Then $D$ also contains
$a_{i_1},\ldots,a_{i_\beta}$ and $b_0,\ldots,b_\beta$.  By the completeness
of $D$, the set $P$ has at least $(\beta+1){\cdot}|S'|$ edges.  Thus $|S'|
\leq h$ because $|P| \leq p=(\beta+1){\cdot}h$. This finishes the proof of
Claim~\ref{claim_thm_table_2} and thus that of Theorem~\ref{thm_table_npc}.
\end{proof}

The next two problems are variants of
Problems~\ref{problem_table_table} and
{\ref{problem_table_graph}}.

\namedkaoproblem{Protection of a Table}{
\label{problem_table_table_undirected}
\begin{itemize}
\item Input: 
The suppressed graph $\HH$ of a table $\TT$ whose total graph is
undirected.
\item Output:
A set $P$ consisting of the smallest number of published cells of
$\TT$ such that the table formed by $\TT$ with the cells in $P$
also suppressed is protected.
\end{itemize}}

\kaoproblem{\label{problem_table_graph_undirected}
\begin{itemize}
\item Input: 
A bipartite undirected graph $\HH=(A,B,E)$.
\item Output: 
A set $P$ consisting of the smallest number of undirected edges
between $A$ and $B$ but not in $E$ such that every nonsingleton
connected component of $(A,B,E \cup P)$ is bipartite-complete.
\end{itemize}}

\begin{lemma}
\label{lem_problem_table_undirected_table_graph_equiv}
Problems~\ref{problem_table_table_undirected} and
{\ref{problem_table_graph_undirected}} can be reduced to each
other in linear time.
\end{lemma}
\begin{proof}
The proof is similar to that of
Lemma~\ref{lem_problem_table_table_graph_equiv}.
\end{proof}

\begin{theorem}[\cite{Kao95.augcomp}]
\label{thm_table_graph_undirected} 
Problem~\ref{problem_table_graph_undirected} can be solved in
optimal $O(|\HH|+p)$ time, where $p$ is the output size.
\end{theorem}

\begin{theorem}\label{thm_table_table_undirected} 
Problem~\ref{problem_table_table_undirected} can be solved in
optimal $O(|\HH|+p)$ time, where $p$ is the output size.
\end{theorem}
\begin{proof}
This theorem follows from
Lemma~\ref{lem_problem_table_undirected_table_graph_equiv} and
Theorem~\ref{thm_table_graph_undirected}.
\end{proof}

\section{Discussions}\label{sec_discussion}
The relationship between the data security of $\TT$ and the
connectivity of $\HH$ are summarized and compared below.

\vspace{0.2in}
\begin{center}
\begin{tabular}{|l|l|}
\hline
Levels of Data Security & Degrees of Graph Connectivity
\\
\hline
\hline
all cells & strongly connected, bridge-free
\\
\hline
all rows and columns & strongly connected, bipartite-2-connected
\\
\hline
all sets of $k$ rows or $k$ columns & strongly connected,
bipartite-$(k+1)$-connected
\\
\hline
the whole table & strongly connected, bipartite-complete
\\
\hline
\end{tabular}
\end{center}
\vspace{0.2in}

\begin{lemma}\label{lem_row_column_imply_cell}
Let $R$ be a row or column of $\TT$.  Let $k$ be the smallest
number of row vertices or column vertices in any nonsingleton
connected component of $\HH$.
\begin{enumerate}
\item
If $R$ is protected, then every suppressed cell in $R$ is also
protected.
\item 
If a set of $k$ rows or $k$ columns of $\TT$ is protected, then
every subset of that set is also protected.
\item
If $\TT$ is protected, then every set of $k-1$ rows or $k-1$
columns is also protected.
\end{enumerate}
\end{lemma}

Note that the converses of the above statements are all false.

\begin{proof}
Statements 1 and 2 are straightforward.  Statement 3 follows from
Theorems {\ref{thm_k_row_column_char_all}} and
{\ref{thm_table_char}}.
\end{proof}

\section*{Acknowledgements}
The author is deeply grateful to Dan Gusfield for his help.  The author
wishes to thank the anonymous referees for very helpful and thorough
comments.


\end{document}